\documentclass[twocolumn]{aastex631}

\usepackage{amsmath, bm}

\received{August 4, 2023}
\revised{September 18, 2023}
\accepted{September 26, 2023}

\submitjournal{ApJ}

\shorttitle{Turbulence in G46.8$-$0.3 and G39.2$-$0.3}
\shortauthors{Shanahan et al.}

\begin{document}

\title{Turbulent Structure In Supernova Remnants G46.8$-$0.3 And G39.2$-$0.3 From THOR Polarimetry}

\correspondingauthor{Russell Shanahan}
\email{rpshanah@ucalgary.ca}

\author[0000-0002-3186-8369]{Russell Shanahan}
\affil{University of Calgary, 2500 University Dr. NW, Calgary AB, T2N 1N4, Canada}

\author[0000-0003-2623-2064]{Jeroen M. Stil}
\affil{University of Calgary, 2500 University Dr. NW, Calgary AB, T2N 1N4, Canada}

\author[0000-0001-8800-1793]{Loren Anderson}
\affil{Department of Physics and Astronomy, West Virginia University, Morgantown, WV 26506, USA}

\author[0000-0002-1700-090X]{Henrik Beuther}
\affil{Max Planck Institute for Astronomy, K{\"o}nigstuhl 17, 69117 Heidelberg, Germany}

\author[0000-0002-6622-8396]{Paul Goldsmith}
\affil{Jet Propulsion Laboratory, California Institute of Technology, 4800 Oak Grove Drive, Pasadena, CA 91109, USA}

 \author[0000-0002-0560-3172]{Ralf S. Klessen}
\affil{Universität Heidelberg, Zentrum für Astronomie, Institut für Theoretische Astrophysik, Albert-Ueberle-Straße 2, 69120 Heidelberg, Germany}
\affil{Universität Heidelberg, Interdisziplinäres Zentrum für Wissenschaftliches Rechnen, Im Neuenheimer Feld 205, 69120 Heidelberg, Germany}

\author{Michael Rugel}
\affil{Center for Astrophysics, Harvard \& Smithsonian, 60 Garden St., Cambridge, MA 02138, USA}
\affil{National Radio Astronomy Observatory, 1003 Lopezville Rd, Socorro, NM 87801, USA}

\author[0000-0002-0294-4465]{Juan D. Soler}
\affil{Max Planck Institute for Astronomy, K{\"o}nigstuhl 17, 69117 Heidelberg, Germany}


\begin{abstract}

We present the continued analysis of polarization and Faraday rotation for the supernova remnants (SNRs) G46.8$-$0.3 and G39.2$-$0.3 in $L$-band (1-2 GHz) radio continuum in The HI/OH/Recombination line (THOR) survey.  In this work, we present our investigation of Faraday depth fluctuations from angular scales comparable to the size of the SNRs down to scales less than our $16\arcsec$ beam ($\lesssim 0.7 \text{ pc}$) from Faraday dispersion ($\sigma_\phi$).  From THOR, we find median $\sigma_\phi$ of $15.9 \pm 3.2 \text{ rad m}^{-2}$ for G46.8$-$0.3 and $17.6 \pm 1.6 \text{ rad m}^{-2}$ for G39.2$-$0.3.  When comparing to polarization at $\lambda6$ cm, we find evidence for $\sigma_\phi \gtrsim 30 \text{ rad m}^{-2}$ in localized regions where we detect no polarization in THOR.  We combine Faraday depth dispersion with the rotation measure (RM) structure function (SF) and find evidence for a break in the SF on scales less than the THOR beam.  We estimate the RM SF of the foreground interstellar medium (ISM) using the SF of extra-Galactic radio sources (EGRS) and pulsars to find that the RM fluctuations we measure originate within the SNRs for all but the largest angular scales.

\end{abstract}


\section{Introduction}
\label{sec:intro}

\indent	The Milky Way contains a multi-phase interstellar medium (ISM) \citep{Ferriere2001,Klessen2016}.  One component of the ISM is the magneto-ionic medium, a mixture of interstellar plasma and magnetic fields \citep[see, for example][for a review]{Beck1996,Heiles2012}. The magneto-ionic medium causes linearly polarized radio waves to undergo a rotation of the polarization angle via the Faraday effect. The investigation of Faraday rotation measures of synchrotron emission from pulsars and extragalactic radio sources (EGRSs) allows one to study interstellar magnetic fields and magnetic turbulence in the Milky Way \citep{Han2001, Han2017, Brown2007, VanEck2011, Beck2015, Haverkorn2015, Jaffe2019, Dickey2022}. 

\indent	Supernova remnants (SNRs) are bright extended synchrotron emitting sources in the Galaxy.  Polarimetry of a SNR reveals a blend of foreground and internal Faraday rotation.  The study of Faraday rotation enables one to investigate internal turbulence in the expanding shell, plus a contribution from electron density and magnetic field fluctuations within the Galactic ISM along the line of sight between the SNR and the observer \citep{Velusamy1975, Milne1990, Simonetti1992, Reich2002, Wood2008, HS2010, Reynolds2012, Farnes2017, Bykov2020, Shanahan2022}.  However, separating internal turbulence of the SNR from turbulence in the ISM is challenging. 

\indent	The simplest case of Faraday rotation occurs where the polarization angle, $\chi$, of a linearly polarized wave changes with wavelength, $\lambda$, according to $\delta \chi = \phi \lambda^2$, where the Faraday depth, $\phi$, is defined as
\begin{equation}
\phi = 0.81 \int \left( \frac{n_e}{\text{cm}^{-3}} \right) \left( \frac{B_{\parallel}}{\mu\text{G}} \right) \left( \frac{dl}{\text{pc}} \right) \left[\text{rad m}^{-2}\right].
\end{equation}
\label{eqn:FDog}
Here $n_e$ is the electron density, $B_\parallel$ is the magnetic field component along the line of sight, and $l$ is the path length from the emitting source to the observer \citep{Burn1966,Brentjens2005}.  This integral is evaluated from the source to the observer, where $\phi > 0$ indicates $B_\parallel$ is pointing towards the observer ,and $\phi<0$ when $B_\parallel$ is pointing away. 

\indent	Complex Faraday rotation is a situation in which a source is observed to have multiple Faraday depth components associated with different Faraday rotating regions within the beam.  Complex Faraday rotation can be observed as a nonlinear relation between $\chi$ and $\lambda^2$ and a change in fractional polarization ($\Pi$) with wavelength.  The observed wavelength range, when investigating polarization, is important for understanding the type of Faraday rotation \citep{Farnsworth2011,Sun2015}.  For example, complex Faraday rotation may appear as simple Faraday rotation when observing over a limited wavelength range.  Considering these factors, complex Faraday rotation can be estimated in multiple ways \citep[][ and refences within]{Alger2021}.  For our purposes, we use the method of Stokes $QU$ fitting \citep{Law2011,OSullivan2017}. 

\indent 	\citet{Shanahan2022} presents polarimetry of SNRs G46.8$-$0.3 and G39.2$-$0.3 in which variation of Faraday rotation and polarized intensity is observed on scales of their beam (16\arcsec).  They also observe complex Faraday rotation in these SNRs, which they conclude is evidence of Faraday depth structure internal to the SNRs.  The distribution of Faraday depths for subregions within SNR G39.2$-$0.3 showed a narrow and broad peak which they propose is a detection of polarization originating from the near and far sides of the SNR shell, respectively. 

\indent	In Section \ref{sec:obs} we present a brief overview of the observations.  For a more detailed description of the observations, calibration and imaging we refer to \citet{Beuther2016} and \citet{Shanahan2022}.  In Section \ref{sec:meth} we outline the methods used in our investigation of complex Faraday rotation and magnetic turbulence.  We present our results in Section \ref{sec:res}, where we present Faraday rotation measure structure functions (Section \ref{sec:snrSFs}), maps of Faraday dispersion (Section \ref{sec:sRM-maps}), and foreground Faraday rotation (Section \ref{sec:SFs}).  We present a discussion of potential selection effects and internal Faraday dispersion in Section \ref{sec:dis}.  In Section \ref{sec:conc} we present a summary and conclusions.


\section{Observations}
\label{sec:obs}
\indent	The THOR survey \citep{Beuther2016} at the Karl G. Jansky Very Large Array (VLA) covers the inner Galaxy in the longitude range $14\fdg5 < l <67\fdg4$ and latitude $-1\fdg25 < b < 1\fdg25$ in C configuration in $L$-band (1 - 2 GHz).  The survey includes OH lines, radio recombination lines, the $\lambda$21 cm line of atomic hydrogen, along with the continuum.  The continuum observations consist of 512 channels ranging from $1 - 2$ GHz, where each channel has a frequency width of 2 MHz.  The $\lambda$21 cm line and total intensity continuum were combined with archival data from the VLA Galactic Plane Survey \citep[VGPS;][]{Stil2006} and the single-dish observations at 1.4 GHz by the Effelsberg continuum survey \citep{Reich1990}.  Only C-configuration data exist for continuum polarization, which samples the sky at 1.5 GHz on angular scales ranging from $\sim15\arcsec$ to $\sim5\arcmin$.

\indent	Radio frequency interference (RFI) flagging as well as standard flux and phase calibration of the continuum data is described in \citet{Beuther2016}.  Standard procedures were followed for the polarization calibration and imaging done in CASA as described in \citet{Shanahan2022}.  The calibrated visibilities were averaged into 8 MHz channels and imaged.  These individual images were combined into image cubes for Stokes $I$, $Q$ and $U$, where each 8 MHz channel has noise of $\sim$0.4 mJy beam$^{-1}$.  

\indent	We use the THOR catalog of polarized extra-Galactic sources in the Galactic longitude range of $39^{\circ} < \ell < 52^{\circ}$ presented in \citet{Shanahan2019}, which are compiled of compact sources brighter than 10 mJy beam$^{-1}$ from source lists in \citet{Bihr2016} and \citet{Wang2018}.  The sample in \citet{Shanahan2019} includes additional polarized components of resolved sources, but excludes sources that, upon visual inspection, were considered to be bright diffuse sources.  As part of our analysis we use polarized subregions of SNRs G46.8$-$0.3 and G39.2$-$0.3, where each subregion we analyze is sized to be $\sim$16\arcsec, the approximate synthesized beam at 1.2 GHz \citep{Shanahan2022}.  For more information on the imaging and calibration, we refer to Section 2 in \citet{Shanahan2022}.  


\section{Methods}
\label{sec:meth}

\subsection{Faraday Rotation Measure Synthesis}
\label{sec:RM}
\indent	The analysis of THOR polarization was done by implementing Faraday rotation measure (RM) synthesis, as described in \citet{Brentjens2005}.  We express linear polarization as a complex function of $\lambda$ in terms of the normalized Stokes parameters $q=Q/I$ and $u=U/I$ as $\mathcal{P} = q + iu$, where $i=\sqrt{-1}$.  Through RM synthesis we derive the dimensionless Faraday dispersion function $\tilde{\mathcal{F}}(\phi)$ by the Fourier transform
\begin{equation}
\tilde{\mathcal{F}}(\phi) = \frac{1}{K} \int_{-\infty}^{\infty} \mathcal{P}(\lambda^2) W(\lambda^2) \exp[-2 i \phi \lambda^2] d\lambda^2.
\label{eqn:FD}
\end{equation}
Here, $K$ is the integral of the weight function, $W(\lambda^2)$.  $W(\lambda^2) = 1$ where measurements exist and $W(\lambda^2) = 0$ where there are no measurements, including when $\lambda^2 < 0$.  The Faraday dispersion function is the complex polarized surface brightness per unit of Faraday depth \citep[for details, see][]{Brentjens2005}.  Different rates of Faraday rotation can be blended by integrating over frequency, solid angle or different synchrotron emission regions along a line of sight that results in complex Faraday rotation.  

\indent	The rotation measure spread function (RMSF) is the Fourier transform of $W(\lambda^2)$ and acts as a point-spread function in Faraday depth, where the full width half max (FWHM) is our resolution in Faraday space.  After RFI flagging, we have a Faraday depth resolution of $\sim$103$\text{ rad m}^{-2}$ that varies by $\sim$2$\text{ rad m}^{-2}$ due to individual channel flagging.  


\subsection{Stokes $QU$ Fitting}
\label{sec:QU}

\indent	 We implement Stokes $QU$ fitting through the use of the RMtools \footnote{https://github.com/CIRADA-tools/RM} package \citep{Purcell2020} by fitting a Faraday rotation model to spectra of polarized emission.  For the simplest case of modelling a polarized signal with the presence of Faraday rotation, we use the equation
\begin{equation}
\mathcal{P} = \mathcal{P}_0 \text{exp}[2i \phi \lambda^2],
\end{equation}
where $\mathcal{P}_0 = \Pi_0 \text{exp}[2i\chi_0]$ is the intrinsic amount of polarization of the synchrotron emission, $\Pi_0$ is the intrinsic fractional polarization, and $\chi_0$ is the intrinsic polarization angle. 

\indent	The mixing of emitting and rotating mediums along a line of sight can cause depolarization towards longer wavelengths.  For the case of turbulent magnetic fields, depolarization occurs when many turbulent cells reside within the synthesized beam \citep{Burn1966, Sokoloff1998, OSullivan2012}, 
\begin{equation}
\mathcal{P} =  \mathcal{P}_0 \text{exp}[-2\sigma_{\phi}^2\lambda^4]\text{exp}[2i(\phi \lambda^2)].
\label{eqn:Burn}
\end{equation}
Here, $\sigma_{\phi}$ is the dispersion about the mean $\phi$ across the source, which we refer to as Faraday dispersion.  For the purpose of this work we only use this model, but for a description of other models see \citet{Sokoloff1998}.  Fitting Equation \ref{eqn:Burn} to the data allows us to measure $\sigma_\phi$ as shown in \citet{Shanahan2022}. If the true depolarization is different, we can at least get an approximate value for the Faraday dispersion, which we will discuss in Section \ref{sec:dis}. 


\subsection{Faraday Rotation Measure Structure Functions}
\label{sec:SF}
\indent	Structure functions (SFs) are used to determine structural scales in fluids and magneto-hydrodynamic (MHD) turbulence \citep{Kolmogorov1941, Goldreich1995}.  Faraday rotation measure (or Faraday depth) structure functions are used to probe variance in magnetic field strength and electron density as a way to investigate magneto-ionic turbulence in the ISM \citep{Simonetti1984, Lazio1990, Clegg1992, Haverkorn2004, Stil2011, Livingston2021}.  The variation in $\phi$ on angular scales larger than the beam can be expressed by the second order structure function (SF),
\begin{equation}
D_2(\bm{\theta}) = \frac{1}{N} \sum_i [\phi(\bm{r}) - \phi(\bm{r} + \bm{\theta})]^2_i,
\label{eqn:SF}
\end{equation}
where $N$ is the number of pairs included in the sum.  For our purposes, we only consider the magnitude of the separation where $|\bm{\theta}|=\theta$, and include all pairs with separation that falls within a narrow range around $\theta$. 

\indent	The assumption that the largest scale of magneto-ionic variations is comparable to the driving scale of fluid turbulence is required to relate the largest scale of magneto-ionic turbulence to the angular scale where the slope breaks \citep{Seta2020, Livingston2021}.  

\indent	From turbulence theories presented in \citet{Kolmogorov1941} and \citet{Goldreich1995} a power-law SF is expected, with less power on small scales than large scales.  Averaging within the beam causes high spatial frequency cut off in the power spectrum that affects the amplitude and slope in the SF \citep{Laing2008}. We derive a similar correction for Faraday rotation SFs of diffuse emission in our SNRs.  


\subsection{Simulated Structure Functions}
\label{sec:simSFs}

\indent	For the purposes of simulating a turbulent Faraday screen, we follow a similar procedure for simulations as outlined in \citet{MD2003}.  For our simulations, we generate a two-dimensional image of Faraday depth with structure similar to what is shown in Figure 1 of \citet{MD2003}.  This is done by generating Gaussian random fields with a power-law SF, where $D_2(\theta) \propto \theta^\alpha$.  We assume a power spectrum form $P_F(\bm{k}) \propto k^\gamma$, where $k$ is the wavenumber.  The input power spectral index of the simulation, $\gamma$, is related to the SF index, $\alpha$, by $\gamma = -\alpha - 2$.  

\indent	Although Faraday depth is an integral of $n_e B_\parallel$ along the line of sight (see Equation \ref{eqn:FDog}), only the total Faraday depth affects our simulations.  \citet{Laing2008} and \citet{Tribble1991} also used a two-dimensional distribution of Faraday depth as a starting point. \citet{Simonetti1984} and \citet{Xu2016} derive power-law SFs of RM by integrating $n_e B_\parallel$ through a turbulent medium.  For us, the details of the integral are less important than the spatial structure of Faraday depth.  The simulation represents the total Faraday depth of a turbulent medium in front of a uniform polarized background. 

\begin{figure}[htb!]
\centering
   \centerline{\includegraphics[width=1\linewidth, angle=0]{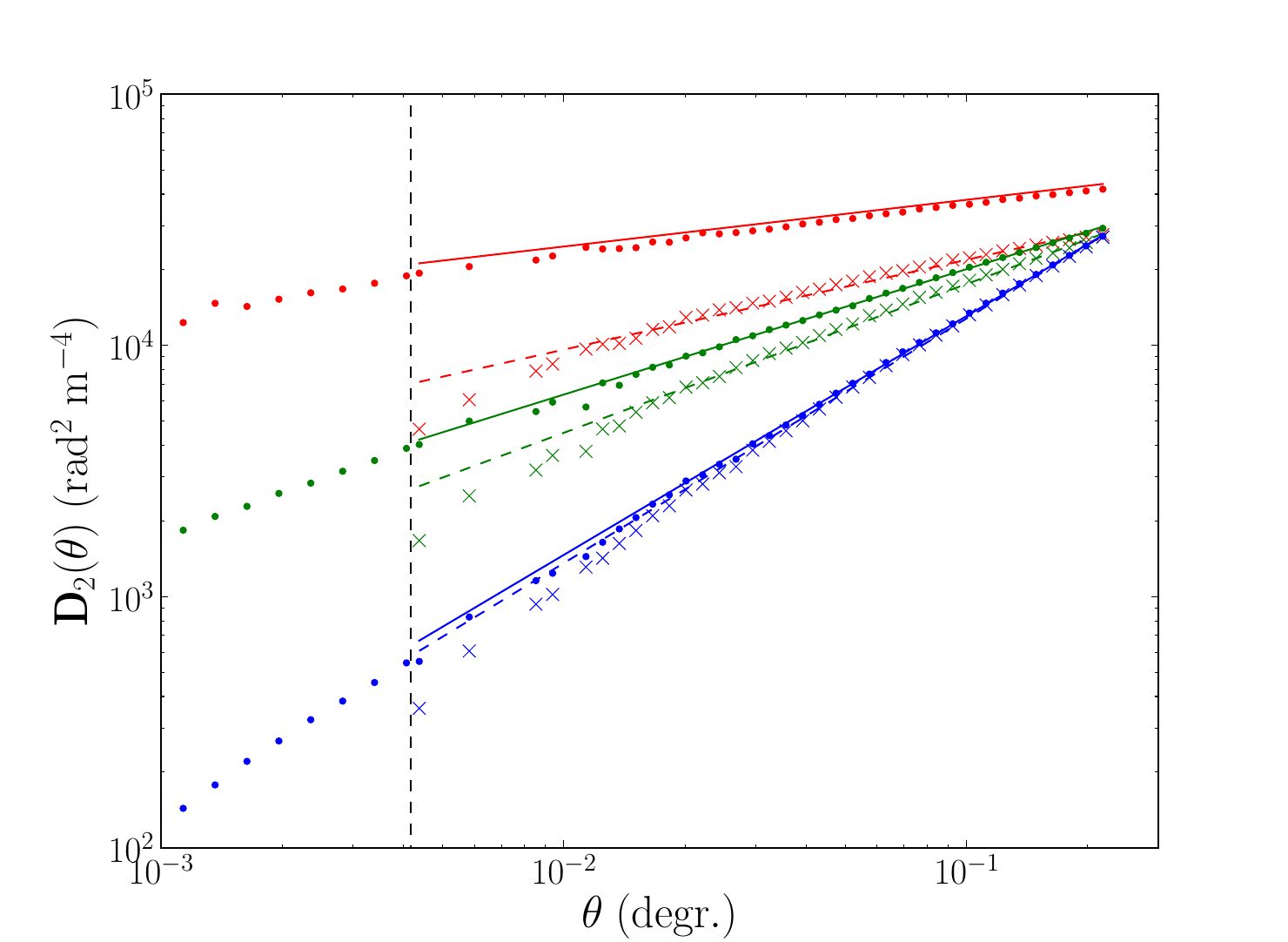}}
   \caption{Simulated RM structure functions.  The red, green, and blue data points correspond to $\gamma = -2.1, -2.5, \text{ and } -3$, respectively.  The dots are SFs from individual pixels.  The crosses are SFs after averaging $Q$ and $U$ over the $16\arcsec$ beam (represented by the vertical black dashed line), as explained in the text.  The colored dashed lines are power-law fits to the crosses.  The solid lines are the dashed lines plus a constant term.  This constant term is the amplitude of each single-pixel SF at $1/3$ the beam. } 
   \label{fig:simSF} 
\end{figure}

\indent	In Figure \ref{fig:simSF} we present SFs from three different simulations where $\gamma = -2.1, -2.5$, and $-3$ for the red, green, and blue data respectively.  The dots are SFs for individual pixels within our simulated image.  The crosses are SFs of RM derived from QU fitting synthetic Q and U spectra that were created by evaluating Stokes $Q$ and $U$ for every pixel as a function of wavelength, assuming a uniform polarized background, and applying a Gaussian weight with a FWHM of $16\arcsec$ (50 pixels) to the $Q$ and $U$ spectra to form a single spectrum for every beam.  We find that the resulting RMs are equivalent with the Gaussian weighted RMs. 

\indent	The SFs using a Gaussian weighted beam have a lower amplitude as well as a higher slope than what we find from the SFs of individual pixels.  In Figure \ref{fig:simSF} we observe that this effect becomes larger as $\gamma \to -2$ (or $\alpha \to 0$).  Fitting a power-law to the individual pixel SFs yields $\alpha=0.195 \pm 0.002, 0.502 \pm 0.004$ and $0.950 \pm 0.004$ for $\gamma = -2.1, -2.5$, and $-3$, respectively.  After beam averaging is applied, the SF becomes curved toward small scales and fitting to a power-law is problematic.  In Figure \ref{fig:simSF} we notice that the slope gets larger while the amplitude gets smaller. \citet{Laing2008} find a similar relation for SFs before and after Gaussian convolution, where, as the SF becomes flatter, the effect of beam averging becomes more pronounced in both amplitude and slope.  

\indent	In our simulations, the SFs derived from individual pixels can be interpreted as what would be found for point sources.  The SFs we find from applying a Gaussian weight to synthesized $Q$ and $U$ spectra correspond with diffuse emission, where Gaussian weighted averaging occurs within the synthesized beam.  From this physical interpretation, we are presented with a problem where the SFs of diffuse emission will misrepresent the true SF of the source.  

\indent	To account for the effect of beam averaging on a SF, we impose a correction to the SFs of diffuse emission sources.  Since we cannot directly measure structures smaller than our beam, we must use simulations to find an approximate correction for unresolved structure.  As shown in Figure \ref{fig:simSF}, this effect is approximately the same as subtracting power on scales $\sim$$1/3$ the beam from the SF.  This happens because scales $\lesssim$$1/3$ the beam will effectively average out in the convolution.  This power ends up as Faraday dispersion within the beam of our observations.  In order to correct for this effect and relate complex Faraday rotation to structure on larger angular scales, we match the observed SF of a SNR to the SF from a simulation with beam averging.  

\section{Results}
\label{sec:res}


\subsection{RM Structure Functions}
\label{sec:snrSFs}

\begin{figure*}[htb!]
\gridline{\fig{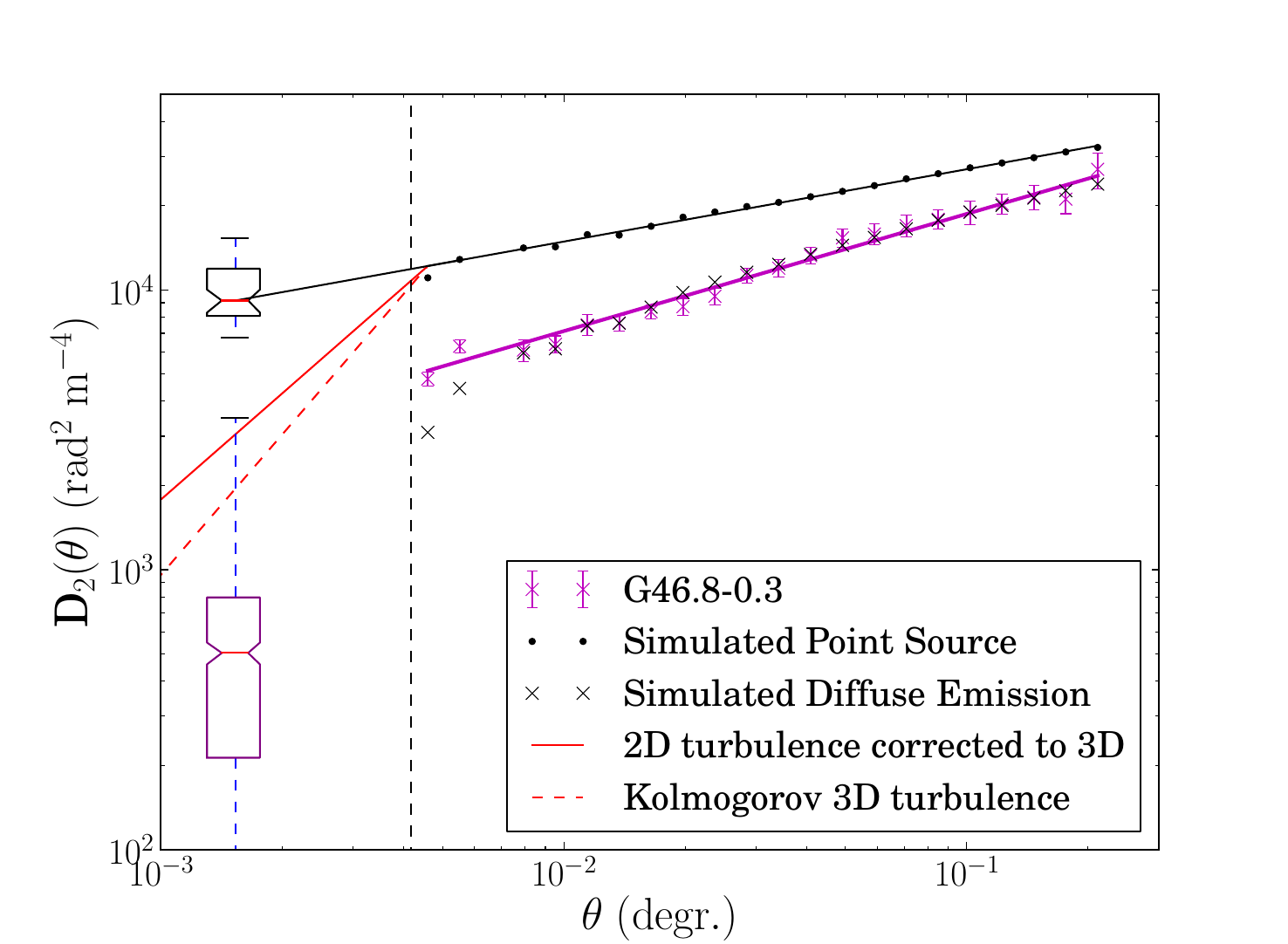}{0.5\textwidth}{(a)}
          \fig{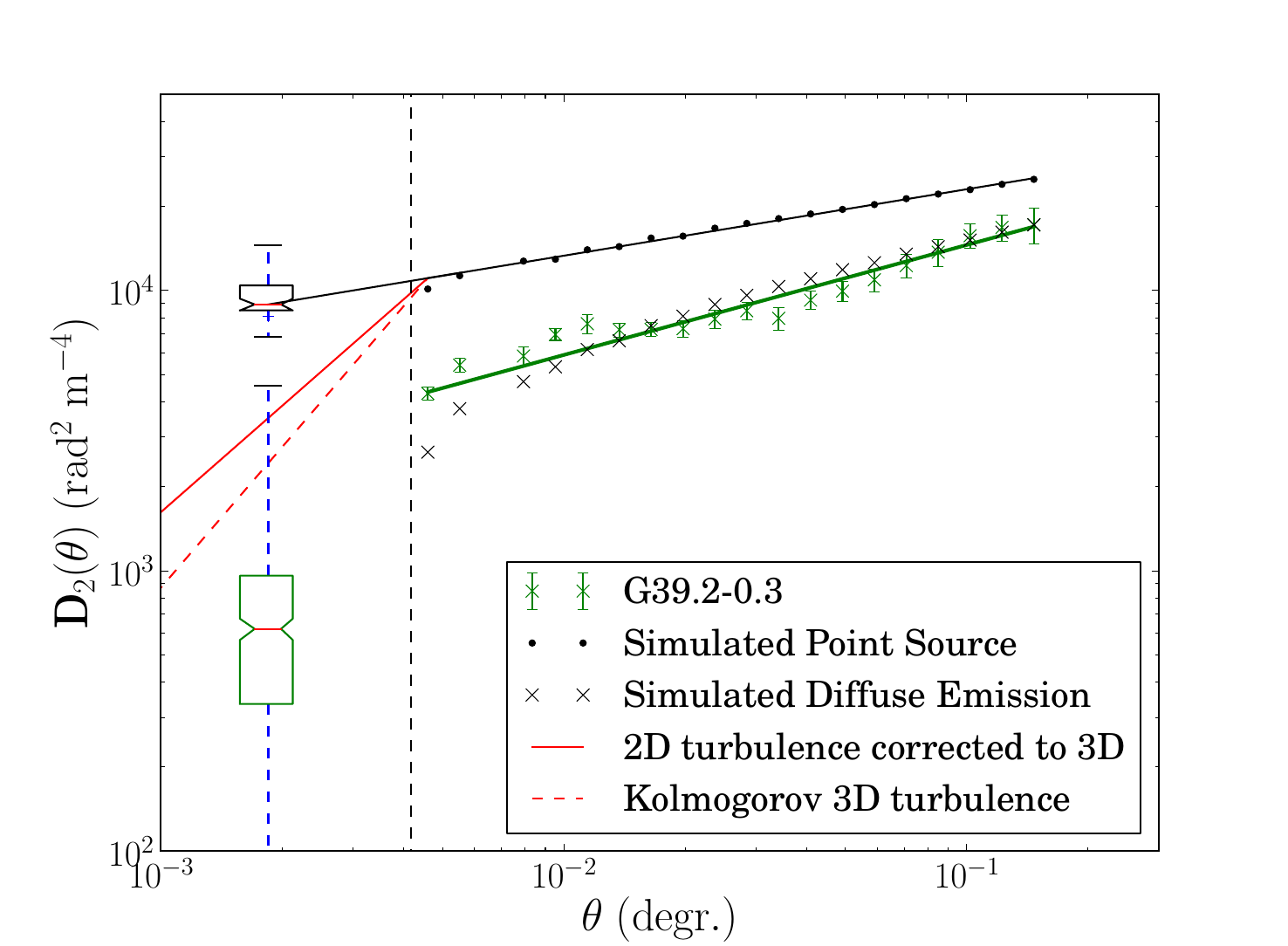}{0.5\textwidth}{(b)}}
\caption{Observed and corrected SFs for SNRs G46.8$-$0.3 \textbf{(a)} and G39.2$-$0.3 \textbf{(b)}.  The colored solid line is a power-law fit to the SNR SF.  The black solid line is a power-law fit to the simulated point source (individual pixel) SF, which has been extended to where it equals the median of the simulation box plot.  The box plots are generated from the distribution of 2$\sigma_\phi^2$ for the SNR (colored) and the simulation (black).  The red line in the box plots are the median of each 2$\sigma_\phi^2$ distribution, the bottom and top of the box are the first and third quartiles, respectively, and the whiskers extend to the maximum and minimum values in the distribution.  The vertical dashed line represents the THOR beam of 16\arcsec. The structure functions are discussed in Section \ref{sec:snrSFs} and the box plots are discussed in Section \ref{sec:sRM-maps}. } 
\label{fig:snrSF}
\end{figure*}

\indent	In Figures \ref{fig:snrSF} (a) and (b) we present the simulated and observed RM structure functions for G46.8$-$0.3 and G39.2$-$0.3, respectively.  The colored crosses in Figures \ref{fig:snrSF} (a) and (b) are the SFs derived from the Faraday depths presented in Figures 3 and 18 of \citet{Shanahan2022}. The black crosses are SFs derived from simulations, where a $16\arcsec$ beam is applied.  The black dots are the SF derived from an individual pixel at the center of the beam in our simulations, which we associate with the SF for point sources.  The colored solid lines in Figure \ref{fig:snrSF} are power-law fits to the SNR RM structure function.   We know that the SNR SFs are affected by beam averaging; therefore, by simulating a beam averaged SF that matches the SNR SF, we can derive a SF for the SNR that is without the effect of beam averaging.  

\indent	The scaling of the simulated SF is done in two steps. First the angular scale of a pixel is determined by equating the simulated beam to the resolution of our data.  Second, the amplitude of the simulated SF with beam averaging is scaled to match the amplitude of the SNR SF.  The scaling accounts for the initially arbitrary scaling of the power spectrum in the simulations.  The same scaling factor is applied to the corrected (single pixel) SF and to the simulation of the complexity in the following subsection.  

\indent	For SNR G46.8$-$0.3, we find that the observed RM structure function has a power-law index, $\alpha = 0.42 \pm 0.01$, and for the corrected SF from simulations,  $\alpha = 0.22 \pm 0.01$.  For SNR G39.2$-$0.3, we find the observed RM structure function has $\alpha = 0.39 \pm 0.01$, and for the corrected SF from simulations,  $\alpha = 0.16 \pm 0.01$.  We see how the effects of beam averaging causes the SFs for each SNR to have a steeper slope and lower amplitude.  The SFs corrected for beam averaging may still contain contributions from plasma anywhere along the line of sight from within the SNR to the observer.  The contribution of the plasma between the SNR and the observer will be estimated in Section \ref{sec:dis}. 

\subsection{Complex Faraday Rotation}
\label{sec:sRM-maps}

\indent	RM fluctuations from beam to beam observed in \citet{Shanahan2022} is the smallest angular scale in our RM structure functions that we can measure directly.  RM fluctuations smaller than the beam manifest themselves as complex Faraday rotation, and the power missing in the beam-averaged SF ends up in Faraday dispersion.  Realizing that the complexity includes RM structure related to the full range of scales smaller than the beam, we combine the measured complexity with the structure function.  

\begin{figure*}[htb!]
\gridline{\fig{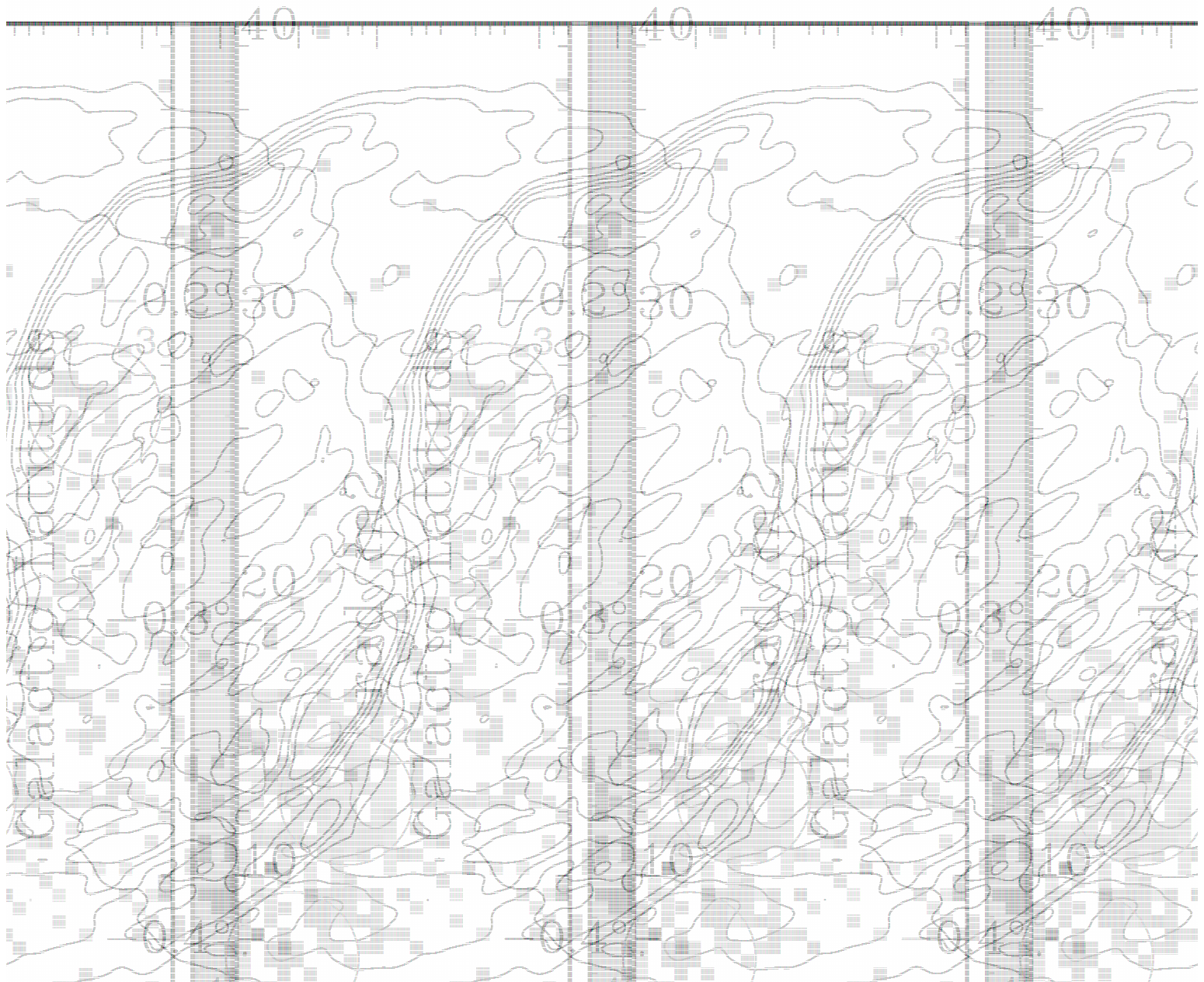}{0.5\textwidth}{(a)}
          \fig{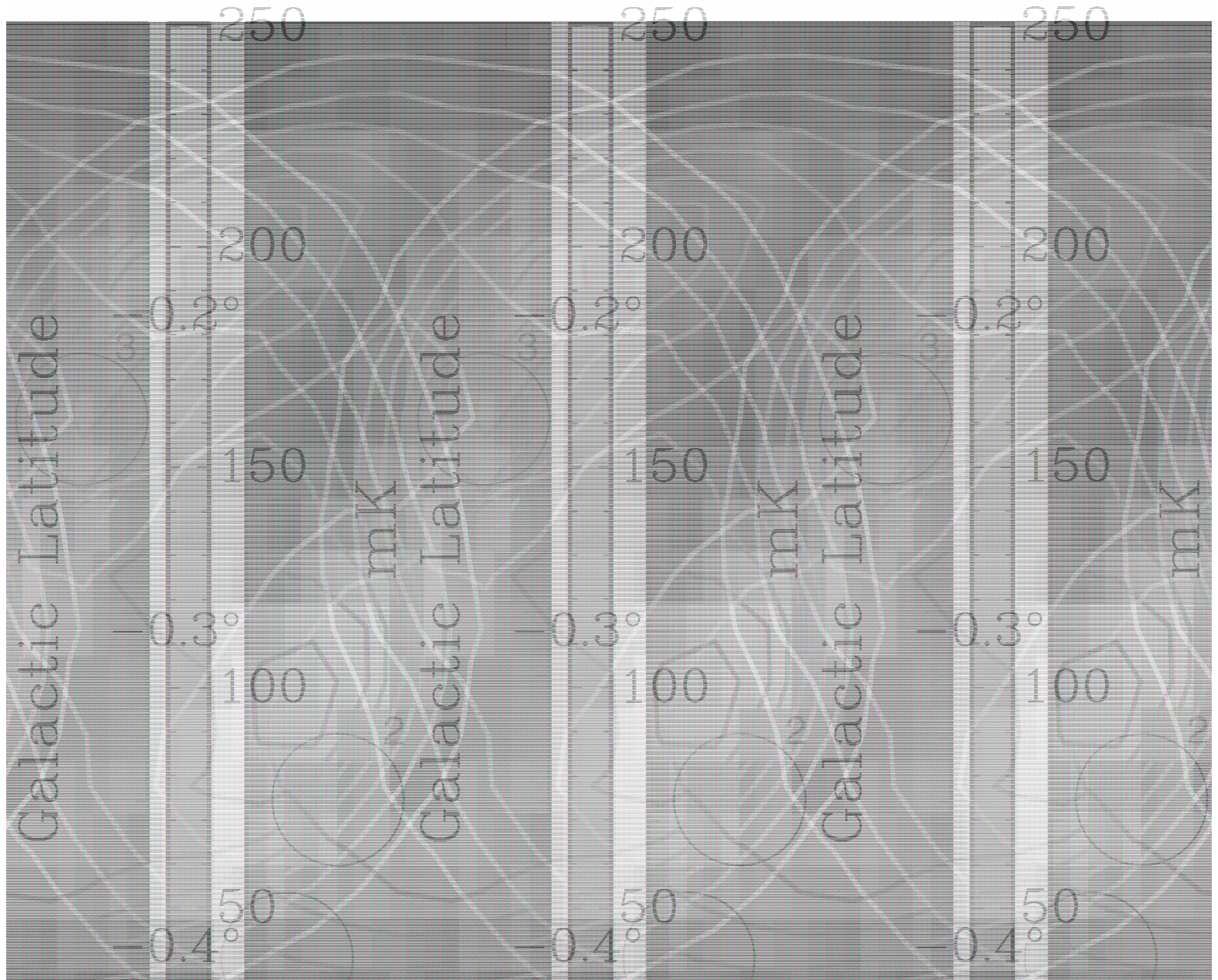}{0.5\textwidth}{(b)}}
\caption{\textbf{(a)} Map of $\sigma_{\phi}$ for SNR G46.8$-$0.3.  Contour levels are 14, 19, 24, 29, 34 and 39 mJy/beam from THOR+VGPS Stoke $I$ at $\lambda$21 cm. Subregions with a purple border indicate two-component Faraday rotation from \citet{Shanahan2022} and are not included when $QU$ fitting using a model of single-component Faraday rotation with Burn depolarization.  \textbf{(b)} Effelsberg polarization intensity map of G46.8$-$0.3 at $\lambda$6 cm with Stokes $I$ contours at 0.2, 0.4, 0.6, 0.8, 1.0, 1.2 and 1.4 K.  The red circles identify regions where we compare THOR polarization to Effelsberg polarization.  The size of the red circles correspond to the Effelsberg beam of 2.5\arcmin. The coordinates of each region are given in Table \ref{tab:metrics}} 
\label{fig:G46}
\end{figure*}

\begin{figure*}[htb!]
\gridline{\fig{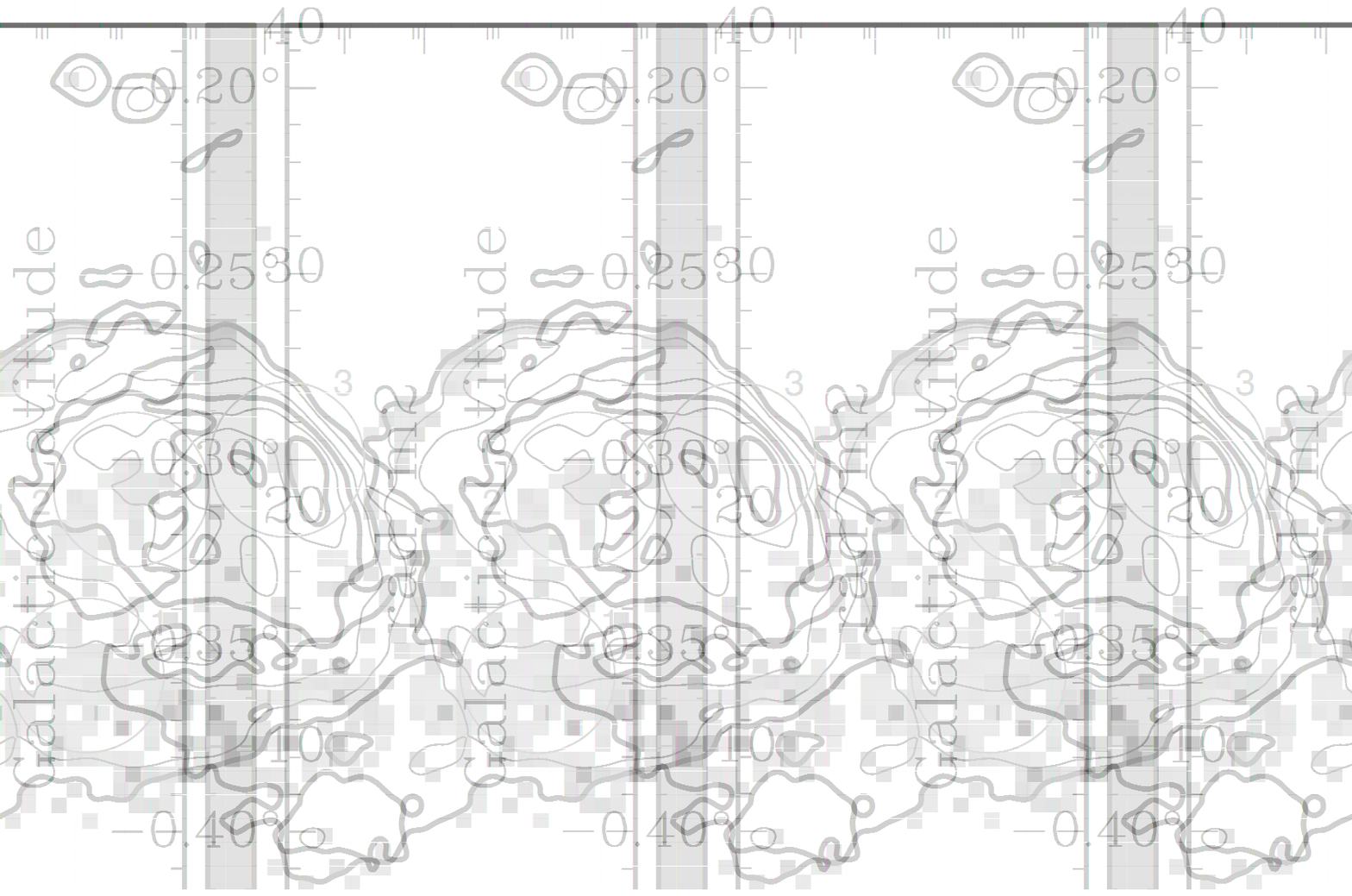}{0.5\textwidth}{(a)}
          \fig{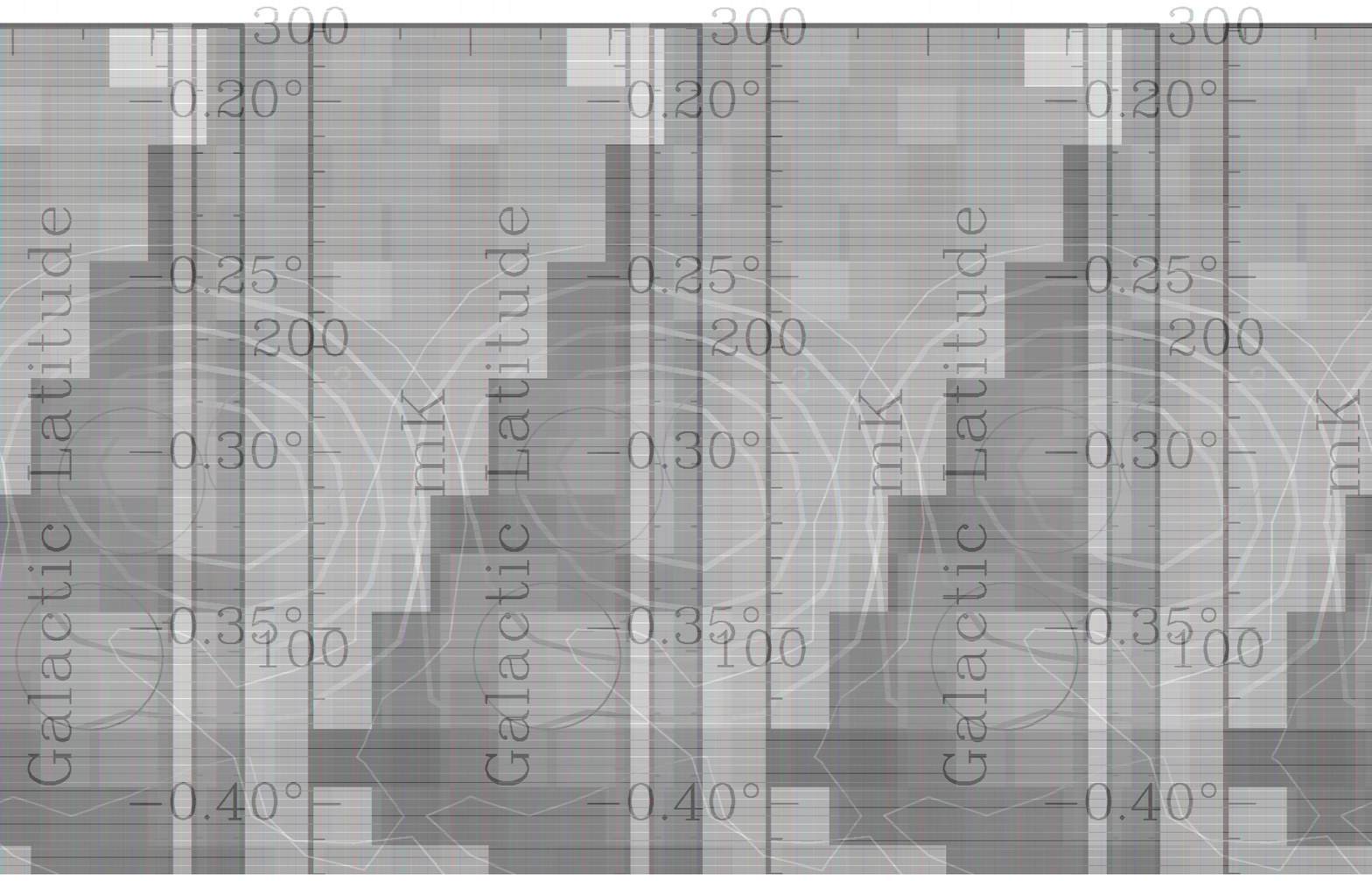}{0.5\textwidth}{(b)}}
\caption{\textbf{(a)} Map of $\sigma_{\phi}$ for SNR G39.2$-$0.3.  Contour levels are 20, 30, 55, 80, 110 and 140 mJy/beam from THOR+VGPS Stoke $I$ at $\lambda$21 cm. Subregions with a purple border indicate two-component Faraday rotation from \citet{Shanahan2022} and are not included when $QU$ fitting using a model of single-component Faraday rotation with Burn depolarization.  \textbf{(b)} Effelsberg polarization intensity map of G39.2$-$0.3 at $\lambda$6 cm with Stokes $I$ contours at 0.5, 1.0, 2.0, 3.0 and 4.0 K.  The red circles identify regions where we compare THOR polarization to Effelsberg polarization.  The size of the red circles correspond to the Effelsberg beam of 2.5\arcmin. The coordinates of each region are given in Table \ref{tab:metrics}} 
\label{fig:G39}
\end{figure*}

\indent	In Figures \ref{fig:G46} (a) and \ref{fig:G39} (a) we present our results of $QU$ fitting as $\sigma_{\phi}$ maps for SNRs G46.8$-$0.3 and G39.2$-$0.3, respectively.  Each colored subregion represents a location where single component Faraday rotation is observed \citep{Shanahan2022}.  The blank subregions with a purple border indicate where two-component Faraday rotation is observed \citep{Shanahan2022}, and are excluded in our $\sigma_{\phi}$ maps because we find a poor fit to Equation \ref{eqn:Burn}.  

\indent	In order to relate our simulations with $\sigma_\phi$ derived from the data, we use the second moment of the Faraday depth distribution.  Complex Faraday rotation can be defined as the second moment of the Faraday depth distribution, 
\begin{equation}
M_2 = \sqrt{ J^{-1} \sum_{i=1}^{N} (\phi_i - \langle \phi \rangle )^2 G_i },
\end{equation}
where,
\begin{equation}
\langle \phi \rangle = J^{-1} \sum_{i=1}^{N} \phi_i G_i, 
\end{equation}
\citep{Livingston2021}.  Here, $\phi_i$ is a Faraday depth component of the form expressed in Equation \ref{eqn:FD}, $G_i$ is the Gaussian weight function at the location of $\phi_i$, and $J$ is the integral over the beam.  

\indent	We derive $\sigma_\phi$ from our simulations by generating a Stokes $Q$ and $U$ spectrum across the THOR $\lambda^2$ range.  This is achieved by adding $\phi \lambda^2$ to the polarization angle, where $\phi$ is the pixel value in the simulation.  We do this for each individual pixel within our beam out to 2$\times$FWHM.  Each beam gives us an integrated Stokes $Q$ and $U$ spectrum to which we apply $QU$ fitting in the same $\lambda^2$ range as the data.  These integrated $Q$ and $U$ spectra include complex Faraday rotation derived from the Faraday depth structure down to scales smaller than the beam.  The distribution of Faraday depth is different for every beam and its statistical properties are similar to the models of \citet{Tribble1991} with a power-law SF.  We use RMtools to fit Equation \ref{eqn:Burn} to the synthesized spectra and we retrieve the RM and $\sigma_\phi$.  We find that $M_2$ and $\sigma_\phi$ from $QU$ fitting beam-averaged $Q$ and $U$ data are equivalent. 

\indent	The Faraday depth dispersion is the combined effect of all structure within the beam with a range of angular scales.  Considering that the SF is a rising function with increasing angular scale, and that the largest angular scale is the size of the SNR, we may identify a characteristic angular scale for the fluctuations contributing to $\sigma_\phi$.  To this end, we associate the value $2\sigma_\phi^2$ from the simulation with the SF amplitude within the beam, and identify the angular scale where this amplitude is achieved in the SF from the simulations.  This occurs at an angular scale of $\theta=5.5\arcsec$ for G46.8$-$0.3 and $\theta=6.7\arcsec$ for G39.2$-$0.3.  We apply the same angular scale to the value $2\sigma_\phi^2$ of the observations of each SNR.  The complete set of $\sigma_\phi$ measurements in a SNR is visualized in Figure \ref{fig:snrSF} in the form of a box plot that marks the median, the lower and upper quartile, as well as the full range of the distribution.  

\indent	The black box plots present the distribution of $2\sigma_\phi^2$ from our simulations.  The $\sigma_\phi$ values from the simulations were derived before any scaling was applied, we then apply the same scaling factor for the simulated SFs to the simulated $2\sigma_\phi^2$ values. From this we find both black box plots have a median $\sigma_\phi \approx 67 \text{ rad m}^{-2}$.  Polarized emission with such a Faraday dispersion would depolarize in $L$-band, but we rely on the ability to scale the depolarization solution according to $\lambda^2 \propto 1/\sigma_\phi$ as explained in \citet{Tribble1991}.   

\indent	The characteristic angular scale associated with the Faraday dispersion is approximately 1/3 of the beam, depending on the slope of the SF.  This is similar to the scale of the power in the SF that is subtracted by beam averaging (see Figure \ref{fig:simSF}).  We find that beam averaging takes away some power from the beam-to-beam variations that ends up as Faraday depth dispersion within the beam.  The amount of power involved corresponds to fluctuations on scales of $\sim 1/3$ the beam.  In this way we connect Faraday dispersion with the fluctuations on larger scales.  

\indent	G46.8$-$0.3 has a median $\sigma_{\phi}$ of 15.9 rad m$^{-2}$ with a median error of 3.2 rad m$^{-2}$ and a standard deviation of 7.4 rad m$^{-2}$.  We find a local maximum of $\sigma_\phi = 22.9 \pm 2.1 \text{ rad m}^{-2}$ at $(\ell,b)=(46\fdg783, -0\fdg353)$, which is within a region where we observe the highest density of detections in Figure \ref{fig:G46} (a).  Some of the strongest signal is observed within this region \citep{Shanahan2022}, yet we find that $\sigma_\phi$ is higher here than in the surrounding areas.  One may notice that $\sigma_\phi$ decreases towards the edge of the SNR, with a local minimum of $\sigma_\phi = 2.0 \pm 1.5$ at $(\ell,b)=(46\fdg805, -0\fdg403)$.  We observe no counterpart to this structure in Faraday depth or fractional polarization \citep{Shanahan2022}, as well as no correlation between $\sigma_{\phi}$ and Stokes $I$ brightness at $\lambda$21 cm. 

\indent	Figure \ref{fig:G46} (b) is a $\lambda$6 cm Effelsberg map of polarized intensity with Stokes $I$ contours (W. Reich private communication) with a beam of 2.5\arcmin.  In the brightest region of G46.8$-$0.3 at $(\ell,b)=(46\fdg666, -0\fdg329)$, we observe a fractional polarization of $\sim18\%$ at $\lambda$6 cm and upper limits to fractional polarization $<1\%$ at $\lambda$21 cm \citep{Shanahan2022}.  From these quantities we derive a minimum $\sigma_\phi$ of 27.4 rad m$^{-2}$.  However, potential beam depolarization in the Effelsberg map implies that $\sigma_\phi$ could be larger.  More about this in Section \ref{sec:dis}.  

\indent	From Figure \ref{fig:G39} (a) we find the distribution of $\sigma_{\phi}$ in G39.2$-$0.3 to be relatively uniform with no obvious counterpart in Faraday depth or fractional polarization to deviations from the median $\sigma_{\phi}$ of 17.6 rad m$^{-2}$.  Subregions within the second lowest contour in Figure \ref{fig:G39} have a median error in $\sigma_{\phi}$ of 1.6 rad m$^{-2}$ with a standard deviation of 7.6 rad m$^{-2}$.  Subregions outside the second lowest contour have a median error in $\sigma_{\phi}$ of 2.4 rad m$^{-2}$ with a standard deviation of 6.9 rad m$^{-2}$.  In Figure 22 (b) of \citet{Shanahan2022}, we present an example of $QU$ fitting for a high signal to noise subregion of G39.2$-$0.3.

\indent	In Figure \ref{fig:G39} (b) we present a $\lambda6$cm Effelsberg map of polarized intensity with Stokes $I$ contours (W. Reich private communication).  When comparing polarization at $\lambda21$cm and at $\lambda6$cm, we find the general locations of polarization are in agreement.  However, we observe a lack of polarization at $\lambda21$cm on the right side of the SNR shell where weak polarization is observed at $\lambda6$cm. 

\indent	For G46.8$-$0.3 and G39.2$-$0.3 we identify three locations where we compare $\lambda$6 cm polarization to $\lambda$21 cm polarization (red circles in Figures \ref{fig:G46} and \ref{fig:G39}).  To account for differing resolutions of the two surveys, we compare the weighted average fractional polarization of subregions at $\lambda$21 cm that fall within the Effelsberg beam ($\bar{\Pi}_\text{21cm}$) to the $\lambda$6 cm fractional polarization closest to the beam centre ($\Pi_\text{6cm}$).  We derive Faraday dispersion using
\begin{equation}
\sigma_{\phi,\text{calc}} = \sqrt{ \frac{-1}{2(\lambda_\text{6cm}^4 - \lambda_\text{21cm}^4)} \ln \left( \frac{\Pi_\text{6cm}}{\bar{\Pi}_\text{21cm}} \right) }. 
\label{eqn:sRMcalc}
\end{equation}
Here, $\lambda_\text{6cm} = 0.06\text{ m}$ and $\lambda_\text{21cm} = 0.21\text{ m}$, which is the reference wavelength we obtained from RM-synthesis of THOR polarization.  In Table \ref{tab:metrics} we present a summary of the relevant parameters for this comparison.  We consider $\sigma_{\phi,\text{calc}}$ to be a lower limit estimation of Faraday dispersion for two reasons.  Firstly, due to the larger beam of the Effelsberg maps, beam depolarization could cause the measured $\Pi_\text{6cm}$ to be low.  Secondly, when applying a Gaussian beam of $2.5\arcmin$ to our data, within each beam there are subregions where we detect no polarization (white subregions).  Some of the non-detections are upper limits, implying the mean fractional polarization ($\bar{\Pi}_\text{21cm}$) is lower. 

\begin{deluxetable*}{ccccccccc}
\tablenum{1}
\tablecaption{Faraday Dispersion Metrics \label{tab:metrics}}
\tablewidth{0pt}
\tablehead{
\colhead{SNR} & \colhead{Region} & \colhead{$\ell$} & \colhead{$b$} & \colhead{$\bar{\Pi}_\text{21cm}$} & \colhead{$\Pi_\text{6cm}$} & \colhead{$\sigma_{\phi,\text{calc}}$} & \colhead{$\bar{\sigma}_\phi$} & \colhead{$N_\text{sub}$} \\
 & & \colhead{$(^{\circ})$} & \colhead{$(^{\circ})$} & $(\%)$ & $(\%)$ & $(\text{rad m}^{-2})$ & $(\text{rad m}^{-2})$ &
 }
\decimalcolnumbers
\startdata
& 1 & 46.802 & $-0.406$ & $6.8 \pm 0.4$ & $25.8 \pm 0.1$ & $18.6 \pm 0.4$ & $9.3 \pm 0.5$ & 31\\
G46.8$-$0.3 & 2 & 46.785 & $-0.356$ & $4.6 \pm 0.1$ & $18.1 \pm 0.1$ & $18.9 \pm 0.2$ & $17.1 \pm 2.3$ & 59\\
& 3 & 46.864 & $-0.234$ & $1.9 \pm 0.1$ & $11.0 \pm 0.1$ & $21.3 \pm 0.2$ & $19.2 \pm 3.5$ & 19\\
\hline
& 1 & 39.248 & $-0.358$ & $2.4 \pm 0.1$ & $11.0 \pm 0.3$ & $19.8 \pm 0.4$ & $18.8 \pm 1.4$ & 52\\
G39.2$-$0.3 & 2 & 39.236 & $-0.308$ & $1.6 \pm 0.1$ & $4.7 \pm 0.1$ & $16.5 \pm 0.6$ & $14.7 \pm 0.9$ & 23\\
& 3 & 39.194 & $-0.300$ & $1.3 \pm 0.2$ & $2.6 \pm 0.1$ & $13.4 \pm 1.8$ & $14.5 \pm 0.7$ & 5
\enddata
\tablecomments{We present polarization and Faraday dispersion metrics.  Here, $\bar{\Pi}_\text{21cm}$ is the weighted mean fractional polarization at $\lambda$21 cm in the Effelsberg beam, $\Pi_\text{6cm}$ is the Effelsberg fractional polarization at $\lambda$6 cm, $\sigma_{\phi,\text{calc}}$ is the calculated Faraday dispersion metric using Equation \ref{eqn:sRMcalc}, $\bar{\sigma}_\phi$ is the weighted mean $\sigma_\phi$ from $QU$ fitting subregions within the Effelsberg beam (red circles in Figures \ref{fig:G46}(a) and \ref{fig:G39}), and $N_\text{sub}$ is the number of subregions within the Effelsberg beam.  It should noted that $\sigma_{\phi,\text{calc}}$ are lower limits due to potential depolarization within the Effelsberg beam.  }
\label{tab:metrics}
\end{deluxetable*}


\subsection{Foreground Faraday Rotation}
\label{sec:SFs}

\begin{figure}[htb!]
\centering
   \centerline{\includegraphics[width=1\linewidth, angle=0]{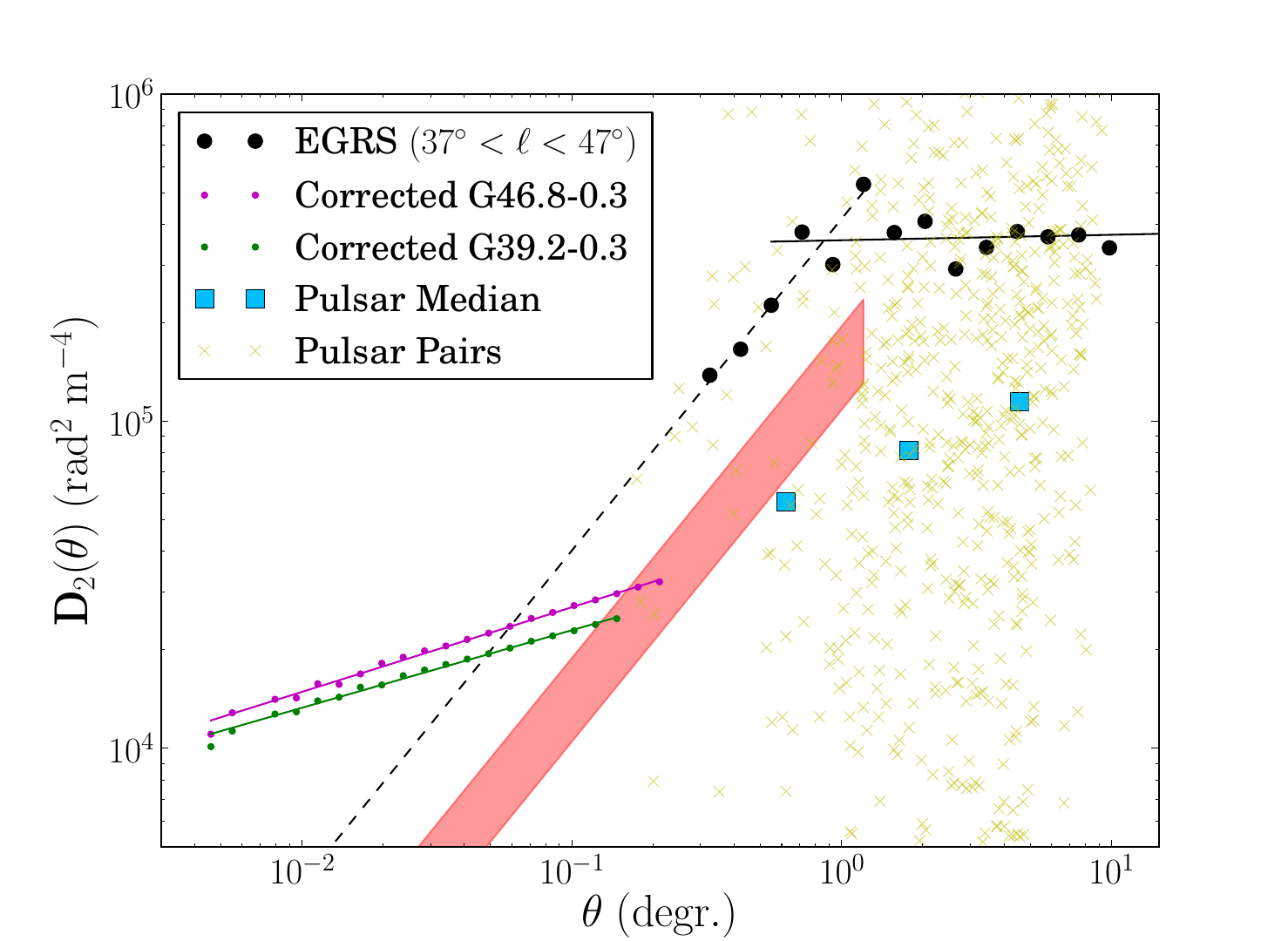}}
   \caption{Structure function for EGRSs and the corrected SNR structure functions.  The solid and dashed black lines are power-law fits to EGRS at $\theta > 0.57^\circ$ and $\theta < 1.25^\circ$, respectively.  The purple and green solid lines are power-law fits to the corrected SFs of G46.8$-$0.3 and G39.2$-$0.3, respectively.  The shaded red region is the approximate SF for Galactic Faraday rotation in the foreground of the SNRs.  The yellow crosses are derived from the ATNF pulsar catalogue \citep{Manchester2005}.  Each cross represents a pair of pulsars with an angular separation of $\theta$ and an amplitude derived with Equation \ref{eqn:SF} with $N=1$.  The blue squares, from left to right, indicate the median values for angular scales $0.1^\circ < \theta \leq 1^\circ$, $1^\circ < \theta \leq 3^\circ$ and $3^\circ < \theta \leq 10^\circ$. } 
   \label{fig:SF} 
\end{figure}

\indent	In Figure \ref{fig:SF} we present the Faraday depth structure function for EGRS within $37^\circ < \ell < 47^\circ$ along with the SNRs G46.8$-$0.3 and G39.2$-$0.3 after correcting for beam averaging (see Section \ref{sec:simSFs}).  The corrected power-law index for G46.8$-$0.3 and G39.2$-$0.3 are $\alpha = 0.22 \pm 0.01$ and $\alpha = 0.16 \pm 0.01$, respectively.  The SF for EGRS has a break in the power-law index at $\theta \approx 1^\circ$, where we find $\alpha = 1.01 \pm 0.25$ for angular scales $\lesssim1.25^\circ$ and $\alpha = 0.02 \pm 0.08$ for angular scales $\gtrsim0.5^\circ$.  

\indent	To estimate which fraction of the EGRS structure function is contaminated by the foreground ISM, we use the ratio of the SNR distances presented in \citet{Lee2020} and distance between the observer and the edge of the Milky Way.  This distance is derived using using a radius of 16.2 kpc for the Milky Way disk and the distance to the Galactic centre of 8.2 kpc \citep{Goodwin1998, Abuter2019}.  These ratios for G46.8$-$0.3 and G39.2$-$0.3 are 3.8 and 2.1, respectively.  We divide the fit to the EGRS SF by this ratio in order to give an approximation of the Galactic RM structure function for the foreground of the SNRs.  This approximate SF is presented in Figure \ref{fig:SF} as the red region where the upper and lower bounds are the EGRS SF divided by the ratios for each SNR.  We find that this approximate foreground EGRS SF has a lower amplitude, and that the largest angular scale of the SNR SFs falls within this range.  

\indent	Another way to constrain the SF of the foreground ISM is by using pulsar data derived from the ATNF pulsar catalogue\footnote{https://www.atnf.csiro.au/research/pulsar/psrcat/} \citep{Manchester2005}.  For our pulsar sample, we subtract the fit of RM versus dispersion measure (DM) from every pulsar RM to remove the contribution of the large scale magnetic field to the difference in RM between two pulsars.  This does not completely eliminate the complications related to the fact that all pulsars in our sample are at different distances, but we use the pulsar RM distances as one way to constrain the small scale RM structure of the ISM between us and the SNRs.  The yellow crosses in Figure \ref{fig:SF} represent $(\text{RM}_i-\text{RM}_j)^2$ for pulsar pairs after subtracting the RM-DM fit, representing individual terms in Equation \ref{eqn:SF}.  The blue squares in Figure \ref{fig:SF} are the median values of the yellow crosses in three bins.   

\indent	We find that the approximate pulsar SF has a lower amplitude and a steeper slope than what is observed for the EGRS SF on similar angular scales.  The median pulsar distance of our sample is 4.8 kpc, which, at an angular separation of $1^\circ$, translates into a distance of 83.8 pc.  This value agrees with the approximate maximum scale for energy injection of turbulence in the Milky Way \citep{Gaensler2005, Haverkorn2006}.  We expect the pulsars to probe somewhat smaller scales than the EGRSs and find that the SF of EGRSs to be flatter than that of pulsars.  Here we are mainly interested in using pulsars as an additional constraint to the SF of the foreground ISM, where the pulsar data give a lower limit to the foreground SF.


\section{Discussion}
\label{sec:dis}

\indent	In this paper we connect the complex Faraday rotation to the rotation measure structure function on larger scales in order to describe the properties in the magnetized plasma of the SNRs over the largest possible range of scales.  We find that Faraday dispersion indicates a small range of Faraday depths on scales smaller than the beam than what is extrapolated from the rotation measure SF.  We discuss potential selection effects related to $QU$-fitting that may arise by deriving $\sigma_\phi$ by fitting Burn depolarization (see Equation \ref{eqn:Burn}) to the data and possible selection effects related to our $\lambda^2$ coverage before discussing possible interpretation. 


\subsection{Selection Effects Of $\sigma_\phi$}
\label{sec:seleff}

\indent	In order to understand how selection effects could bias $\sigma_\phi$, we first must question whether the Burn depolarization model is adequate.  A consequence of \citet{Tribble1991} is that Burn depolarization underestimates $\sigma_\phi$ at longer wavelengths.  \citet{Tribble1991} demonstrates that wavelength dependent depolarization is not as extreme as the Burn model suggests and that the Burn model is only a good approximation when the scale of fluctuations is much less than the beam.  \citet{Tribble1991} suggest that if $\sigma_\phi \lambda^2 < 1$ the Burn depolarization model provides a good approximation of $\sigma_\phi$.  If $\sigma_\phi \lambda^2 > 1$ Burn depolarization underestimates the true $\sigma_\phi$ of the source because a given amount of depolarization can be achieved with a smaller $\sigma_\phi$.  We fit Equation \ref{eqn:Burn} to our data because the model is readily available in the $QU$-fitting code.  

\begin{figure}[htb!]
\centering
   \centerline{\includegraphics[width=1\linewidth, angle=0]{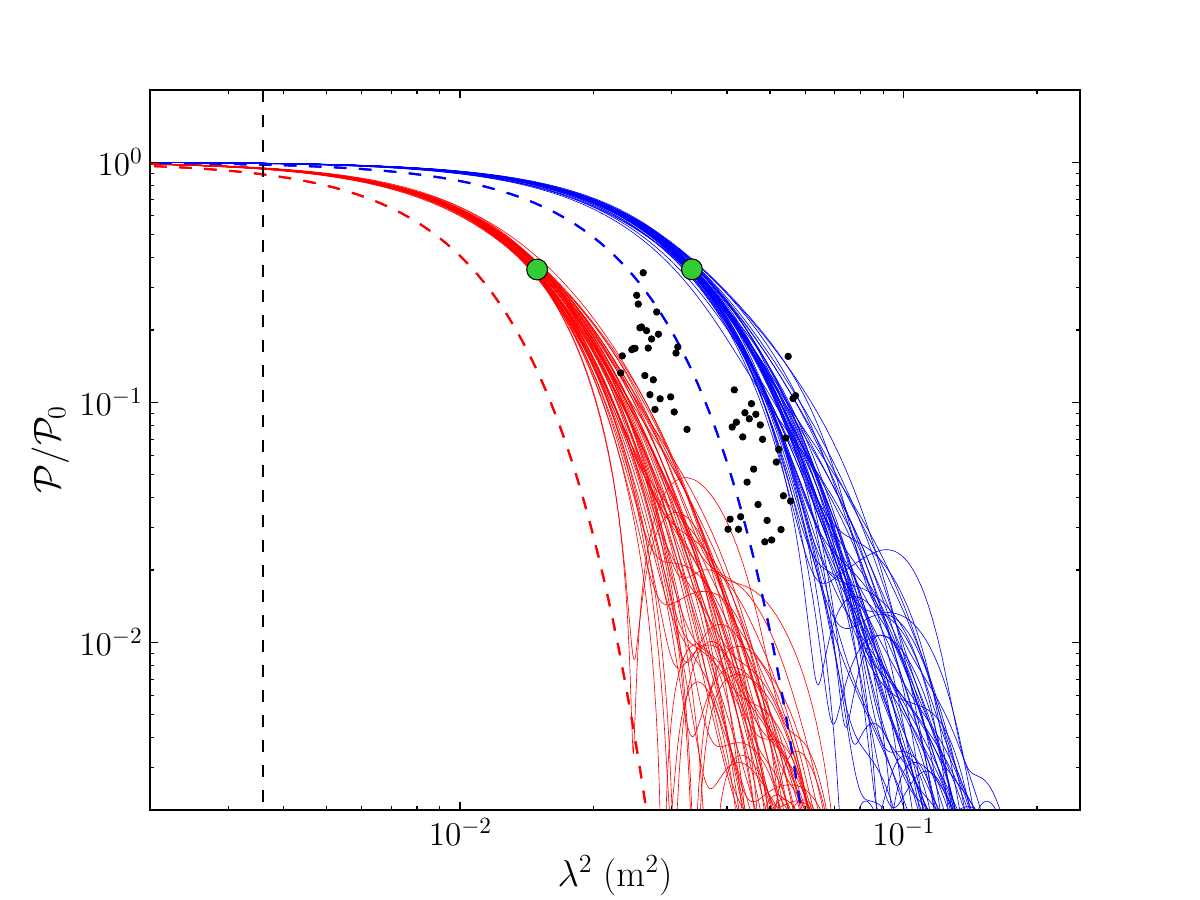}}
   \caption{Depolarization across wavelengths expected from turbulence cells within the synthetized beam according to the modelling in \citet{Burn1966} and \citet{Tribble1991} (see Section \ref{sec:seleff} for details). The red curves represent Tribble (solid) and Burn (dashed) for $\sigma_\phi = 67 \text{ rad m}^{-2}$.  The blue curves represent Tribble (solid) and Burn (dashed) for $\sigma_\phi = 30 \text{ rad m}^{-2}$.  The green dots indicate where $\sigma_\phi \lambda^2 = 1$ in each case.  The black dots are data from the subregion at $(\ell,b)=(39\fdg238, -0\fdg372)$, for which we find $\sigma_\phi = 42.6 \pm 5.3 \text{ rad m}^{-2}$.  The vertical black dashed line shows where $\lambda = 6 \text{ cm}$.  } 
   \label{fig:Tribble} 
\end{figure}

\indent	In Figure \ref{fig:Tribble} we present a comparison of Burn and Tribble depolarization in the context of our SNRs.  The solid lines represent Tribble depolarization of a power law structure function, where each individual line comes from an independent simulation.  For the Tribble model we define $\sigma_\phi$ as the standard deviation of Faraday depth in the half power radius of the beam.  In order to observe the spread in Tribble depolarization at longer wavelengths, the simulation was repeated 54 times for each value of $\sigma_\phi$.  

\indent	As in \citet{Tribble1991}, we see that the Burn model leads to stronger depolarization at the same wavelength.  The range of depolarization factors for the Tribble model increases with wavelength, because every beam represents a random selection of turbulent structure in RM.  This leads to an important concept that a range of $\sigma_\phi$ is to be expected when applying Burn depolarization to a partially resolved turbulent plasma with a single power-law RM SF. 

\indent	The black dots in Figure \ref{fig:Tribble} are data from $(\ell,b)=(39\fdg238, -0\fdg372)$ in Region 1 of G39.2$-$0.3 with $\sigma_\phi = 42.6 \pm 5.3 \text{ rad m}^{-2}$.  We plot $\Pi_\lambda/\Pi_\text{6cm}$, where $\Pi_\lambda$ is the fractional polarization derived from THOR at wavelength $\lambda$ and $\Pi_\text{6cm}$ is the fractional polarization observed from Effelsberg at $\lambda6$ cm.  At longer wavelengths we observe a relatively flat spectrum and an uptick at shorter wavelengths.  We observe similar behavior in various Tribble simulations where the curve either flattens or increases at longer wavelengths.  The data show an example that is not well fitted by Burn depolarization. 

\indent	For both G46.8$-$0.2 and G39.2$-$0.3 we find $\sigma_\phi \approx 30 \text{ rad m}^{-2}$ to be on the high side of what we detect with the Burn model.  For $\sigma_\phi = 30 \text{ rad m}^{-2}$, the point where $\sigma_\phi \lambda^2 = 1$ occurs at $\lambda=18.3 \text{ cm}$, which falls roughly in the middle of the THOR continuum wavelength range (see Figure \ref{fig:Tribble}).  If there are regions where $\sigma_\phi \gtrsim 30 \text{ rad m}^{-2}$ we would not detect them, because of the strong depolarization in our wavelength range.  However, we may occasionally detect significantly higher values of $\sigma_\phi$ as in $(\ell,b)=(39\fdg238, -0\fdg372)$, because of the sporadic extended wing of the Tribble depolarization curves.  There could be higher values of $\sigma_\phi$ in the SNRs, but we cannot justify $\sigma_\phi$ as high as $67 \text{ rad m}^{-2}$ because the SNRs would be completely depolarized in $L$-band (see Figure \ref{fig:Tribble}).  However, it is possible we are missing polarization in the THOR data in regions where polarization is observed with Effelsberg.  

\indent	Figure 8 of \citet{Patnaik1990} presents a polarization map at $\lambda$6 cm of G39.2$-$0.3 with a resolution of 25\arcsec, which is similar to the $16\arcsec$ beam of THOR.  We find that regions where polarized emission is observed at $\lambda$6 cm in \citet{Patnaik1990} agree with the locations where we detect polarized emission at $\lambda$21 cm.  Figure 8 of \citet{Patnaik1990} does show more polarization at $\lambda6$ cm in the top right part of the SNR shell where Stokes $I$ is brightest (Region 3 in Figure \ref{fig:G39}).  In this region we detect polarized emission as well as fractional polarization upper limits $<1\%$ \citep[see Figure 23 of][]{Shanahan2022}.  \citet{Patnaik1990} present a fractional polarization of $\sim6\%$, which we use to derive $\sigma_{\phi,\text{calc}} \approx 22 \text{ rad m}^{-2}$.  We find this to be significantly higher than $\sigma_{\phi,\text{calc}}$ using Effelsberg data (see Table \ref{tab:metrics}).  In the lower right side of the SNR shell of G46.8$-$0.3 we find a similar region where polarization is detected at $\lambda6$ cm that is depolarized at $\lambda21$ cm.  However, the comparison is not as robust due to the large difference in angular resolution between THOR and Effelsberg observations.  Nonetheless, with these considerations it is possible that we only detect polarization in areas that happen to have $\sigma_\phi \lesssim 30 \text{ rad m}^{-2}$.  

\indent	The corrected SFs presented in Figure \ref{fig:snrSF} suggest much higher fluctuations than the observed Faraday dispersion.  While $\sigma_{\phi,\text{calc}}$ technically provides a lower limit to $\sigma_\phi$, it is interesting that $\bar{\sigma}_\phi$ and $\sigma_{\phi,\text{calc}}$ in Table \ref{tab:metrics} tend to agree, with the exception of Region 1 in G46.8$-$0.3.  In locations where we have a higher density of detections, such as Region 1 in G46.8$-$0.3, RM fluctuations might be locally lower than the global average that is indicated by the SF.  There could be higher values of $\sigma_\phi$ elsewhere in the supernova remnants, but we cannot justify $\sigma_\phi$ being as high as the extrapolation of corrected SF suggests.  

\indent	We conclude that the observed Faraday dispersion is lower than the extrapolation from the corrected SF for regions where we detect polarization.


\subsection{Internal Faraday Dispersion Of SNRs}

\indent	\citet{Shanahan2022} provide an analysis of Faraday rotation and polarization of G46.8$-$0.3 and G39.2$-$0.3 in $L$-band from the THOR survey.   They observe two-component complex Faraday rotation, which they argue is evidence for internal Faraday rotation within the SNRs that can be separated as signal from the near and far sides of the SNR shell.  This internal Faraday rotation is of the order of $\sim150 \text{ rad m}^{-2}$.  

\indent	With the considerations provided in Section \ref{sec:seleff}, we can justify that the global depolarization for both SNRs is not as severe as the extrapolation of our corrected SF suggests.  We also acknowledge that the distribution of the observed values of $\sigma_\phi$ is lower than what we expect when comparing THOR polarization to $\lambda6$ cm.  Therefore, the distribution of $\sigma_\phi$ must fall somewhere between the observed and simulated box plots in Figure \ref{fig:snrSF}.  The observed complexity associated with the fiducial scale we derive indicates a slope that is comparable within the range of uncertainty to the transition from 2D turbulence corrected to 3D and Kolmogorov 3D turbulence. There are other models of turbulence that predict different slopes \citep[for examples see,][]{Bec2007, Falgarone2015}.  Our data do not allow to discriminate between different turbulent models.  Even when considering selection bias in our data, we propose that the observed distribution of Faraday dispersion provides evidence for a break in the power-law of the SFs for G46.8$-$0.3 and G39.2$-$0.3 at scales less than our beam. 

\indent	In Figure \ref{fig:SF} we present an approximate SF for Galactic Faraday rotation in the foreground of the SNRs as a red region.  At the largest angular scales of G46.8$-$0.3 and G39.2$-$0.3, we find the amplitudes of the SNR SFs agree with our estimation of the foreground ISM Faraday depth structure.  This is an indication that the RM structure at large angular scales of the SNRs could be affected by Faraday rotation in the foreground ISM.  It is evident that the SFs for G46.8$-$0.3 and G39.2$-$0.3 deviate in both slope and amplitude from our foreground estimation for Galactic Faraday rotation at smaller angular scales.  This deviation suggests that the SFs of the SNRs originate from internal Faraday depth structure as opposed to foreground turbulence in the ISM and that the power on smaller scales is dominated by internal turbulence of the SNR.


\section{Summary and Conclusions}
\label{sec:conc}

\indent	In this work we present a continuation of the study of polarization and Faraday rotation of SNRs G46.8$-$0.3 and G39.2$-$0.3 presented in \citet{Shanahan2022}.  Here we investigate turbulence in these SNRs over a range of scales traced by Faraday rotation. 

\indent	By simulating turbulent Faraday screens with various SF slopes, we demonstrate how beam averaging affects the resulting SF in concordance with previous results by \citet{Laing2008}.  We find that SFs without beam averaging have a flatter slope as well as a higher amplitude than SFs where Gaussian weighted beam averaging is applied.  We correct for this effect through the use of simulations. 

\indent	We use Stokes $QU$ fitting to derive the Faraday dispersion for each subregion where single-component Faraday rotation is observed in \citet{Shanahan2022}.  We observe a median $\sigma_\phi$ of $15.9 \pm 3.2 \text{ rad m}^{-2}$ and $17.6 \pm 1.6 \text{ rad m}^{-2}$ for G46.8$-$0.3 and G39.2$-$0.3, respectively. 

\indent	We use Effelsberg $\lambda6$ cm polarization as a way to verify and extend our measurements of $\sigma_\phi$.  We found evidence for $\sigma_\phi \gtrsim 30 \text{ rad m}^{-2}$ in localized regions where we detect no polarization in THOR.  Following previous modelling of Faraday dispersion, we assume the Faraday dispersion derived from Stokes $QU$ fitting to be a lower limit to the true Faraday dispersion.  However, the comparison to polarization at $\lambda6$ cm suggests only a marginal increase in $\sigma_\phi$ from what we detect from THOR.  

\indent	We use simulations to compare the corrected SNR SF to the observed Faraday depth dispersion.  Even when considering selection bias in the observed $\sigma_\phi$, we find evidence for a break in the RM SF on scales smaller than our beam ($16\arcsec$).  

\indent	We provide an approximate SF for foreground Galactic Faraday rotation from which the SNR SFs deviate at scales $\lesssim 0.5^\circ$.  We attribute this deviation as an indication of Faraday depth structure which is internal to the SNR as opposed to originating from the Galactic foreground.  


\begin{acknowledgments}

The authors thank the anonymous referee for constructive comments that helped make the paper clearer.  The National Radio Astronomy Observatory is a facility of the National Science Foundation operated under cooperative agreement by Associated Universities, Inc. The authors acknowledge the use of the RMtools package written by Cormac Purcell.  The authors acknowledges J.-F. Robitaille for simulation code.  J.M.S. acknowledges the support of the Natural Sciences and Engineering Research Council of Canada (NSERC), 2019-04848.  R.S.K. acknowledges funding from the European Research Council via the ERC Synergy Grant ``ECOGAL'' (project ID 855130), from the German Excellence Strategy via the Heidelberg Cluster of Excellence (EXC 2181 - 390900948) ``STRUCTURES'', and from the German Ministry for Economic Affairs and Climate Action in project ``MAINN'' (funding ID 50OO2206). The team in Heidelberg also thank for computing resources provided by {\em The L\"{a}nd} and DFG through grant INST 35/1134-1 FUGG and for data storage at SDS@hd through grant INST 35/1314-1 FUGG.   J.D.S. acknowledges funding by the European Research Council via the ERC Synergy Grant ``ECOGAL -- Understanding our Galactic ecosystem: From the disk of the Milky Way to the formation sites of stars and planets'' (project ID 855130).  M.R. is a Jansky Fellow of the National Radio Astronomy Observatory.  This research was carried out in part at the Jet Propulsion Laboratory, which is operated by the California Institute of Technology under a contract with the National Aeronautics and Space Administration (80NM0018D0004). 

\end{acknowledgments}


\bibliography{refs}{}
\bibliographystyle{aasjournal}

\end{document}